\documentclass{iau_FM}
\usepackage{graphicx}
\usepackage{hyperref}

\title[The rotation profile of the hotter Solar
atmosphere] %% give here short title %%
{Exploring the variation in the dynamic rotation profile of the hotter solar atmosphere using mutliwavelength data}

\author[S. Routh et. al]   %% give here short author list %%
{Srinjana Routh$^{1,2}$,
Bibhuti Kumar Jha$^3$,
Dibya Kirti Mishra$^{1,2}$,
Tom Van Doorsselaere$^4$,
Vaibhav Pant$^1$,
Subhamoy Chatterjee$^3$
 \and Dipankar Banerjee$^{1,5,6}$}

\affiliation{$^1$Aryabhatta Research Institute of Observational Sciences, \\Nainital-263002, Uttarakhand, India\\email: {\tt srinjana@aries.res.in}\\[\affilskip]
$^2$Mahatma Jyotiba Phule Rohilkhand University, \\Bareilly-243006, Uttar Pradesh, India \\[\affilskip]
$^3$Southwest Research Institute, Boulder, CO 80302, USA\\[\affilskip]
$^4$Centre for mathematical Plasma Astrophysics, Mathematics Department, KU Leuven, Celestijnenlaan 200B bus 2400, B-3001 Leuven, Belgium\\[\affilskip]
$^5$Indian Institute of Astrophysics, \\Koramangala, Bangalore 560034, India\\[\affilskip]
$^6$Center of Excellence in Space Sciences India, IISER Kolkata,\\ Mohanpur 741246, West Bengal, India
}

\pubyear{2024}
\jname{Astronomy in Focus, Focus Meeting 8} 
\editors{Alexander G. Kosovichev, ed.}
\begin{document}

\maketitle

\begin{abstract}
The global rotational profile of the solar atmosphere and its variation at different layers, although crucial for a comprehensive understanding of the dynamics of the solar magnetic field, has been a subject to contradictory results throughout the past century. In this study, we thereby unify the results for different parts of the multi-thermal Solar atmosphere by utilizing 13 years of data in 7 wavelength channels of the Atmospheric Imaging Assembly (AIA) atop the Solar Dynamic Observatory (SDO). Using the method of image correlation, we find that the solar atmosphere exhibits a rotational profile that is up to 4.18\%  and 1.92\% faster at the equator and comparatively less differential than that of the photosphere, as derived from Doppler measurements and sunspots, respectively and exhibits variation at different respective heights. Additionally, we find results suggestive of the role played by the rooting of different magnetic field structures on a comparison with helioseismology data. 
%% add here a maximum of 10 keywords, to be taken form the file <Keywords.txt>
\end{abstract}

\firstsection % if your document starts with a section,
              % remove some space above using this command.
\section{Introduction}

Solar differential rotation has intrigued scientists since its 17\textsuperscript{th}-century discovery. Early studies using sunspot tracking measured photospheric differential rotation, expressed by the equation,
\begin{equation} 
\Omega= A + B\sin^2{\theta} +C \sin^4{\theta}  \label{eq1} \end{equation} 
 where $\theta$ is latitude, $A$ is the equatorial rotation rate, and $B$ and $C$ are latitudinal gradients. With the development of more sophisticated instruments, measurement techniques like spectroscopy and the advent of the technique of helioseismology over the past century, new insights have been gained on the rotation of the solar photosphere and the interior leading to a generalized consensus.

The last century witnessed significant advancements in instruments and measuring techniques, which improved sunspot tracking (\cite[Jha \etal\ 2021]{Jha2021}) and led to measuring solar rotation based on new techniques such as spectroscopy (\cite[Howard \& Harvey 1970]{Howard1970}). The field of helioseismology (\cite[Charbonneau \etal\ 1999]{Charbonneau1999}) emerged, allowing exploration of the rotational profile of the Solar interior, providing insights into the possibility of the Solar magnetic field originating from a region called the tachocline (\cite[Antia \etal\ 1998]{Antia1998}). Consequently, extensive research in this field over the last few decades has established a consensus on the differential rotation profile of the Sun in both its interior and photosphere.

Nonetheless, many questions remain, particularly regarding the higher solar atmosphere's rotational profile, where magnetic fields dominate dynamics. Early studies by (\cite[Livingston 1969]{Livingston1969}) found the chromosphere rotates faster than the photosphere, while other recent studies focusing on different parts of the solar atmosphere have often arrived at a contradiction (\cite[Bertello \etal\ 2020, Mishra \etal\ 2024]{Bertello2020,Mishra2024}). In an attempt to explain the faster rotation of the Solar Atmosphere, \cite[Weber (1969)]{Weber1969} proposed a theory involving the transport of angular momentum by the magnetic field to higher heights from the photosphere. Recent studies by \cite[Li \etal\ (2019)]{Li2019} have emphasized the role of small-scale magnetic structures in understanding the faster rotation of the Solar Atmosphere. However, the reason why certain studies seem to find the solar atmosphere rotating slower was explored by \cite[Badalyan (2010)]{Badalyan2010}, who proposed a bi-modal nature in the rotational profile of the Solar Atmosphere with Solar activity cycle and increasing heliocentric distances. 

Despite all these studies utilizing various methods and data sets, a comprehensive understanding of the global rotational profile of the solar atmosphere above the photosphere and how it varies
across different layers remains elusive due to the diverse results
obtained. In an attempt to address this gap, this study adopts a
more focused approach by utilizing a single tracer-independent
method, that is, the method of image correlation, to analyze the
extensive data set provided by the Atmospheric Imaging
Assembly (AIA) of the solar Dynamic Observatory (SDO)
from the period of 2010–2023.

\section{Data \& Methodology}
The Atmospheric Imaging Assembly (\cite[AIA; Lemen \etal\ 2012]{Lemen2012}) atop the Solar Dynamic Observatory (\cite[SDO; Chamberlin \etal\ 2012]{Chamberlin2012}) has been providing data since its launch in 2010 till present. The AIA provides data in wavelength bands sensitive to primary ions Fe {\sc xviii} ($94$\,{\AA}), Fe {\sc viii, xxi} ($131$\,{\AA}), Fe {\sc ix} ($171$\,{\AA}), Fe {\sc xii, xxiv} ($193$\,{\AA}), Fe {\sc xiv} ($211$\,{\AA}), He {\sc ii} ($304$\,{\AA}), and Fe {\sc xvi} ($335$\,{\AA}), thereby probing the solar atmosphere at different temperatures ranging from $\approx 10^4$ K to $10^7$ K. In addition, one of the telescopes of the AIA observes in C {\sc iv} line near $1600$\,{\AA} and the nearby continuum at $1700$\,{\AA} as well as in the visible continuum at $4500$\,{\AA}.

\begin{figure}[!h]
% \vspace*{-2.0 cm}
\begin{center}
 \includegraphics[scale=0.22]{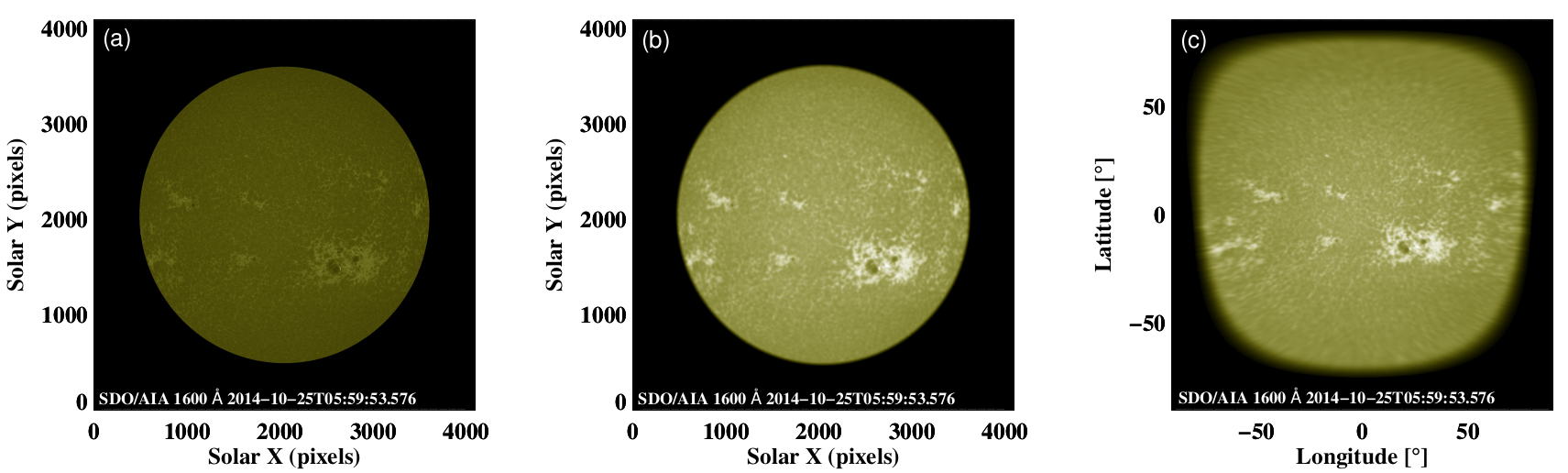} 
 \caption{(a) A Level 1.5 image in the $1600$\,{\AA} wavelength band. (b) The same image after Gaussian smoothing operation is performed. (c) The image after conversion to heliographic coordinates.}
   \label{fig1}
\end{center}
\end{figure}

In this study we utilize the data for 7 wavelength bands, 6 Extreme Ultraviolet ($131$\,{\AA}), $171$\,{\AA}), $193$\,{\AA}, $211$\,{\AA}, $304$\,{\AA}, and $335$\,{\AA}) and 1 Ultraviolet ($1600$\,{\AA}) from 2010 May till 2023 August at a cadence of 6 hours\footnote{AIA data can be downloaded from \href{http://jsoc.stanford.edu/ajax/exportdata.html}{here}}. This choice of cadence makes sure that only longer living features contribute to the correlation analysis without developing much. After these level 1 images were converted to Level 1.5 {\it aia\_prep.pro} procedure available under AIA/SolarSoft (\cite[Freeland \& Handy 1998]{Freeland1998}) in the Interactive Data Language (IDL), smoothing operation with a gaussian kernel was performed on them to minimize the effects of small-scale features even further (\autoref{fig1} (b)). Following these all these images were subjected to conversion to heliographic coordinates (\cite[Thompson 2006]{Thompson2006}) with a resolution of $0.1^{\circ}$/pixel (\autoref{fig1} (c)). Following this, two consecutive images were then divided into latitudinal bins, $15^{\circ}$ in extent, extending till $\pm 60^{\circ}$. These bins were then subjected to the image correlation method, where the cross-correlation coefficient (CC) was calculated for different values of shifts ($\Delta\phi$) between the two images. The shift at which the CC is maximum is used to calculate  the rotation rate $\Omega_{\theta}$\footnote{$\Omega_{\theta}=\frac{\Delta\phi}{\Delta t}$} at the particular latitude $\theta$. An even more detailed discussion of this method is outlined in \cite[Mishra \etal\ (2024)]{Mishra2024} and \cite[Routh \etal\ (2024)]{Routh2024}.

 \vspace*{-0.8 cm}
\section{Results}
\begin{figure}[!ht]
% \vspace*{-2.0 cm}
\begin{center}
 \includegraphics[scale=0.4]{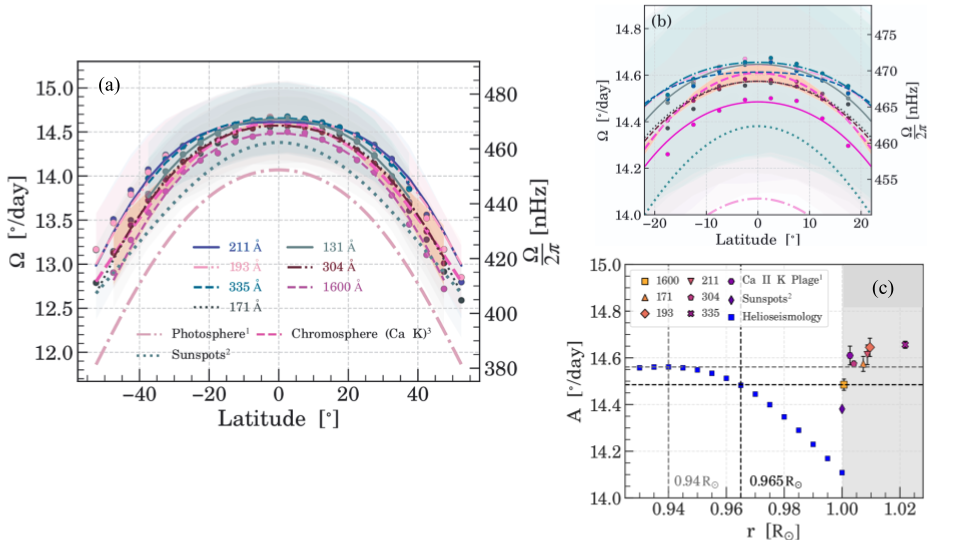} 
% \vspace*{-2.0 cm}
 \caption{(a) All Rotational Profiles of the Solar atmosphere when compared to photspheric plasma ($^1$\cite[Snodgrass 1984]{Snodgrass1984}) sunspots ($^2$\cite[Jha \etal\ 2021]{Jha2021}) and and the chromosphere as obtained from plages of Ca K ($^3$\cite[Mishra \etal\ 2024]{Mishra2024}). (b) A zoomed in version of the equatorial rotation rate near the equatorial regime showing clear variations in the rotation profiles. (c) A variation of equatorial rotation rate $A$ from the interior to the atmosphere.}
   \label{fig2}
\end{center}
\end{figure}

Upon obtaining the individual rotation rates at each latitude for each wavelength band, we compute a temporal average of $\Omega_{\theta}$ weighted by the individual correlation coefficients. This was done after discarding all results corresponding to low values of correlation coefficients (CC$<0.65$ for $131$\,{\AA}, $171$\,{\AA}, $193$\,{\AA}, $211$\,{\AA}, and $335$\,{\AA} and CC $<0.70$ for $304$\,{\AA}, and $1600$\,{\AA}). this was then fit with \autoref{eq1} to obtain the best fit parameters $A, B, C$ for each layer representing different parts of the solar atmosphere. Upon comparing with the parameter we see all these profiles lie well above the rotational profile for photosphere, thereby suggesting that the solar atmosphere does rotate faster and comparatively less differentially (\autoref{fig2} (a)). On looking at the variation near the equatorial regime, we notice there seems to be a variation the the rotation rate (\autoref{fig2} (b)). 

A comparison with the helioseismological data was then performed to present a more complete picture of the variation of the equatorial rotation rate $A$ from the interior, as obtained in helioseismological studies, to the atmosphere, as obtained in this study. Interestingly, we also notice a match in the equatorial rotation rate at the depths of $0.94R_{\odot}, 0.965R_{\odot}$ with $304$\,{\AA}, $171$\,{\AA} and $1600$\,{\AA} (\autoref{fig2} (c)). This can be explained throught the theoretical perspective of \cite[Weber (1969)]{Weber1969}, who suggested deeper rooting of magnetic field lines can help carry angular momentum from a faster rotating interior to the solar atmosphere. A detailed perspective on this has been discussed in \cite[Routh \etal\ 2024]{Routh2024}.

\section{Acknowledgement} 
We thank Council of Scientific \& Industrial Research (CSIR), Government of India for providing the travel grant enabling the in-person participation of this conference in Cape Town, South Africa. We also thank IAU-GA for providing registration fee waiver for attending the conference. We also thank H.M. Antia for providing the helioseismological data used in this study.

\end{document}